\documentclass[a4paper,12pt]{article}

\usepackage{amsmath}
\usepackage{amssymb}
\usepackage{amstext}
\usepackage{ulem}
\usepackage{calc}
\usepackage{vmargin}
\usepackage{graphicx}

\begin{document}
	
	\large
	
	\title{{\bf General solutions to the equations of acoustics in inhomogeneous media and gas dynamics}}	
	\author{ \bf O.V. Kaptsov
		\\ Federal Research Center for Information and Computational Technologies,\\
		Novosibirsk, Russia
		\\ E-mail: profkap@mail.ru}
	
	\date{}
	
	\maketitle
	
	This paper considers one-dimensional equations of acoustics equations of inhomogeneous media and the system of gas dynamics equations with constant entropy. Using the Riemann approach, the gas dynamics equations are reduced to a second-order linear hyperbolic equation with variable coefficients. Solutions to this equation are constructed using Euler–Darboux transformations. This allows us to find new exact solutions of the equations of acoustics and gas dynamics, depending on two arbitrary functions.
	
	
	
	\numberwithin{equation}{section}
	
	\section{Introduction}
	
	In the classical work \cite{Riem}, Riemann introduced characteristic invariants (Riemann invariants) for the system of gas dynamics equations
		\begin{equation} \label{gd}
		\rho_t + (\rho u)_x = 0, \quad u_t + u u_x + p_x/\rho = 0,
	\end{equation}
		where \(\rho\) is the density, \(p\) is the pressure, and \(u\) is the velocity. It is assumed that the pressure is a function of the density. The speed of sound is defined by the formula \(c = \sqrt{p'_\rho}\), and the Riemann invariants have the form
	\[
	r = u + \int \frac{c\,d\rho}{\rho}, \qquad s = u - \int \frac{c\,d\rho}{\rho}.
	\]
	As Riemann showed, in terms of these invariants, the integration of the system (\ref{gd}) reduces to solving a second-order linear partial differential equation
		\begin{equation} \label{Wrs}
		w_{rs} = m \cdot (w_r - w_s),
	\end{equation}
		where \(m\) is a function depending on \(r-s\). For example, if \(p = \rho^\gamma\), then (\ref{Wrs}) is an equation of the Euler–Poisson–Darboux (EPD) type of a special form
	\begin{equation} \label{EPD}
		w_{rs} = \frac{n}{r-s} (w_r - w_s),
	\end{equation}
	where \(n=\frac{\gamma-3}{2(\gamma-1)}\).
	
	As is well known \cite{Mises, Roz}, if \(n\) is an integer, then there exist solutions of equation (\ref{EPD}) of the form
		\begin{equation} \label{eq:1.4}
		w = \sum_{i=0}^k a_i \frac{\partial^i X}{\partial r^i} + \sum_{j=0}^m b_j \frac{\partial^j T}{\partial s^j},
	\end{equation}
	where \(a, b\) are some specific functions of \(r, s\), and \(X, T\) are arbitrary smooth functions of \(r\) and \(s\) respectively. Such solutions, depending on two arbitrary functions, will be called general solutions of rank \((k,m)\) or simply general solutions.
	The construction of general solutions of the type (\ref{eq:1.4}) was studied by Euler \cite{Euler}. Therefore, equations (\ref{Wrs}) and those derived from (\ref{Wrs}) by variable transformations will henceforth be referred to as Riemann–Euler equations.
	It is clear that by a change of variables, equation (\ref{Wrs}) is transformed into the equation
		\begin{equation} \label{vtt}
		v_{tt}=v_{xx} + b(x) v_x.
	\end{equation}
		In turn, the last equation reduces to linear acoustic equations of inhomogeneous media
		 \cite{Br}
		\begin{equation} \label{eq:1.5}
		p_{tt} = f(y) p_{yy}.
	\end{equation}
		The general solution of equation (\ref{eq:1.5}) for \(f = y^{\frac{4m}{2m-1}}\), with integer \(m\), was found by Euler \cite{Euler}. In a recent work \cite{Kaptsov1}, the general solution was obtained for another function \(f\).
	
	In this paper, general solutions of equation (\ref{eq:1.5}) are constructed for some special functions \(f\) and new solutions of the gas dynamics systems (1.1) are presented. The paper has the following structure. In the second section, the Euler–Darboux transformation
		\[
	z = q(v_x + S v),
	\]
		is found, which maps solutions of equation (\ref{vtt}) to solutions of the equation
	\[
	z_{tt} = z_{xx} + g z_x,
	\]
where \(q, S, g\) are some functions of \(x\). This allows us to find general solutions for a series of equations (\ref{vtt}) using the solution of the equation
	\[
	v_{tt} = v_{xx}.
	\]
	In particular, it is proved that the equation
	\[
	v_{tt} = v_{xx} + \frac{(n+2)x^{n+1} - n a}{x(x^{n+1}+a)} v_x 	\]
	has a general solution for  even \(n\) and any $a\in\mathbb{R}$. In the third section, solutions of the equations of acoustics and gas dynamics are constructed. It is shown how to find the general solution of the acoustic equation
	\[
	p_{tt} = (n+1)^{2}(1-a y)^{\frac{2n}{n+1}} y^{\frac{2n+4}{n+1}}p_{yy}
	\]
	for even \(n\) and any \(a\in\mathbb{R}\). For a special equation of state, a solution to the gas dynamics system (1.1) depending on two arbitrary functions has been obtained.
	
	
	\section{Euler–Darboux Transformation}
	
	The most well-known special case of the Riemann–Euler equation is the EPD equation (\ref{EPD}).
		By a change of variables, the EPD equation is reduced to the form
		\begin{equation} \label{epd}
		u_{tt} = u_{xx} + \frac{n}{x} u_x, \quad n \in \mathbb{R}.
	\end{equation}
		It is easy to see \cite{Euler} that the transformations
		\begin{equation} \label{trans}
		v = \frac{u_x}{x}, \qquad v = x u_x + (n-1)u 
	\end{equation}
		map solutions of equation (\ref{epd}) to solutions of the equations
		\[
	v_{tt} = v_{xx} + \frac{n+2}{x} v_x, \qquad v_{tt} = v_{xx} + \frac{n-2}{x} v_x.
	\]
	Since the wave equation
	\[
	u_{tt} = u_{xx}
	\]
	has the solution \(u = T(t+x) + X(t-x)\), where \(T, X\) are arbitrary smooth functions, the transformations (\ref{trans}) yield solutions of equation (\ref{epd}) for any even \(n\).
	
	The Riemann–Euler equation (\ref{Wrs}) is reduced to the equivalent form
		\begin{equation} \label{utt=}
		u_{tt} = u_{xx} + G(x)u_x,
	\end{equation}
		where \(G\) is some function of \(x\). The question arises:  is it possible, using differential substitutions
		\begin{equation} \label{v=}
		v = q(x)(u_x + S(x)u),
	\end{equation}
	to map solutions of equation (\ref{utt=}) to solutions of a similar equation
	\begin{equation} \label{vtt=}
		v_{tt} = v_{xx} + G_1(x)v_x
	\end{equation}
	with some function \(G_1\)?
	
	Substituting (\ref{v=}) into (\ref{vtt=}), we obtain a certain differential equation \(E\). We require that the resulting equation be a consequence of equation (\ref{utt=}). To do this, we express the derivatives \(u_t\), \(u_{tx}\) using the right-hand side of equation (\ref{utt=}) and substitute them into equation \(E\). As a result, we obtain an equation depending on the variables \(u_x\), \(u\), \(u_{xx}\). We require that the coefficients of these variables be equal to zero. Then we obtain three equations:
		\[
	G q - G_1 q - 2 q_x = 0,
	\]
		\begin{equation} \label{qxx}
		q_{xx} + (G_1 + 2S) q_x + (2S' - G S + G_1 S - G') q = 0,
	\end{equation}
		\[
	S q_{xx} + 2 S^{\prime} q_x + S G_1 q_x + q G_1 S^{\prime} + q S''= 0.
	\]
	
	From the first equation we have
		\begin{equation} \label{G1}
		G_1 = G - 2(\ln q)',
	\end{equation}
		Expressing \(q_{xx}\) from the second equation and substituting it into the third, we obtain an equation for the function \(S\):
	\[
	S'' - 2 S' S + (G S)' = 0.
	\]
	Integrating the last equation, we obtain a Riccati equation
		\[
	S' = S^2 - G S + A,
	\]
		where \(A\) is an arbitrary constant. Setting \(S = -(\ln h)'\), we obtain a linear equation for the function \(h\)
				\begin{equation} \label{h2}
		h'' + G h' + A h = 0.
	\end{equation}
	Let us introduce a new function \(r =1/q\). Then equation (\ref{qxx}) takes the form
		\begin{equation} \label{r2}
		r^{\prime\prime} + G r^{\prime} + \left(G^{\prime} + 2 (\ln h)^{\prime\prime}\right)r = 0.
	\end{equation}
		Thus, the following assertion is proved.
	
	\noindent
	\textbf{Lemma 1.} Let \(u\) be a solution of equation (\ref{utt=}). Then the function
		\[
	v = \frac{u_x - (\ln h)' u}{r}
	\]
	satisfies equation (\ref{vtt=}) if
	\begin{equation} \label{G1=}
		G_1 = G + 2 \frac{r'}{r},
	\end{equation}
	and the functions $h$ and $r$ are solutions of equations \eqref{h2} and \eqref{r2} respectively.
	
	\noindent
	\textbf{Proposition 1.} If \(h\) is a solution of equation (\ref{h2}), then the function \(r= (\ln h)'\) satisfies equation (\ref{r2}).
	
	\noindent
	The proof is carried out by direct substitution of \(r = (\ln h)'\) into equation (\ref{r2}) and using equation (\ref{h2}).
	
	\noindent
	\textbf{Proposition 2.} If \(A = 0\), then the general solution of equation (\ref{h2}) has the form
		\[
	h = c_1 + c_2 \int e^{-\int G \, dx} \, dx,
	\]
		where \(c_1, c_2\) are arbitrary constants.
	
	\noindent
	Indeed, if \(A = 0\), then setting \(h' = y\), we obtain a first-order linear equation, which is easily integrated.
	
	Consider as an example equation (\ref{epd}) and suppose that \(A=0\). Then, according to Lemma 1 and Proposition 2, the functions \(h\) and \(r\) are given by 
		\[
	h = c_1 + c_2 x^{1 - n},
	\]
	\[
	r =\frac{c_3 + c_4 \left[ (n - 3) c_1 (c_1 x^{n+1} + c_2 (n + 1) x^2) - c_2^2 (n + 1) x^{3 - n} \right] }{c_1 x^n + c_2 x},
	\]
	where \(c_1, c_2, c_3, c_4\) are arbitrary constants. Moreover, the function \(G_1\) 
	is given by formula (\ref{G1=}).

	

\noindent
\textbf{Proposition 3.}
	Let $u$ be a solution of equation (\ref{utt=}). Then the function $v = u_x / r$ is a solution of equation (\ref{vtt=}), where  $G_1$ is given by (\ref{G1=}) and
\begin{equation} \label{r=}
	r = c_1 e^{-\int G\,dx} + c_2 e^{-\int G\,dx} \int e^{\int G\,dx} dx, \quad c_1, c_2 \in \mathbb{R}.
	\end{equation}	
	\noindent
	\textit{Proof.} Indeed, if \(A = 0\), then one solution of equation (\ref{h2}) is 
	 \(h = 1\). For \(h=1\), equation (\ref{r2}) reduces to a first-order equation	
	\[
	r' + G r = c_1, \quad c_1 \in \mathbb{R}.
	\]
	The last equation has solution (\ref{r=}), and it remains to apply Lemma 1.

For example, consider the equation (\ref{epd}). Then, according to Proposition~3, the function \(r\) has the form
\[
r = c_1 x^{-n} + c_2x, \quad c_1, c_2 \in \mathbb{R},
\]
and the function \(G_1\) is given by the formula
\begin{equation} \label{G11}
	G_1(x) = \frac{c_2 (n + 2) x^{n + 1} - n c_1}{c_1 x + c_2x^{n+2}}.
\end{equation}

	We can again apply Proposition 3. If the function \(G\) in equation (\ref{utt=}) is given by  (\ref{G11}), then the corresponding function \(r\) has the form
		\[
	r = \frac{\left[ c_3 x^n + c_4 \left( c_1 c_2 (n + 3)(n - 1) x^{n +2} + c_2^2 (n -1) x^{2n +3} - c_1^2 (n + 3)x \right) \right]}{(c_1 + c_2 x^{n+1})^2},
	\]
	where \(c_3, c_4\) are arbitrary constants. The function \(G_1\) is found by formula (\ref{G1=}). With each new step, the function \(r\) becomes more cumbersome.
	
	Let us give examples of solutions to the equation
	(\ref{h2}) for \(A = -1\), \(G = n/x\) for \(n = 2, 4, 6\)
		\[
	\begin{aligned}
		&h = c_1 \frac{\sinh x}{x} + c_2 \frac{\cosh x}{x}, \quad n = 2 ,\\
		&h = c_1 \frac{x-1}{x^3}e^{x}  + c_2  \frac{x+1}{x^3}e^{-x}, \quad n = 4 ,\\
		&h = c_1  \frac{x^2 - 3x + 3}{x^5}e^{x} + c_2 \frac{x^2 + 3x + 3}{x^5}e^{-x} , \quad n = 6 .
	\end{aligned}
	\]
		Using Proposition 1, the corresponding functions \(r\) can be found.
	
	Here is another useful assertion that can help in finding solutions.
	\noindent
	\textbf{Lemma 2.} Let \(A_1, \ldots, A_n\) be pairwise distinct constants, and \(h_1, \ldots, h_n\) be the corresponding solutions of equation (\ref{h2}). Then the transformation
	\[
	z = \frac{W(u,h_1, \ldots, h_n)}{W(h_1^{\prime}, \ldots, h_n^{\prime})}
	\]
	maps a solution of equation \eqref{utt=} to a solution of the equation
	\begin{equation} \label{ztt}
	z_{tt} = z_{xx} + 2 z_x \frac{\partial}{\partial x}
	\left( \ln \frac{W(h_1', \ldots, h_n')}{W(h_1, \ldots, h_n)} \right),
	\end{equation}
	where \(W\) denotes the Wronskian of the corresponding functions.
	
	\noindent
	The proof of the lemma is presented in \cite{KaptsovBook}.

	To apply this assertion, consider the wave equation
	\[ 	u_{tt} = u_{xx}, 	\]
	 which has a general solution
	 $$ u=T(t+x) +X(t-x) ,$$
	 where $T$ and $X$ are arbitrary functions.
		 Since \(G = 0\) in this case, the solutions to equation (\ref{h2}) for \(A_i < 0\) are the functions 
	 \[
h_i = c_i \exp(x \sqrt{-A_i}) + b_i \exp(-x \sqrt{-A_i}), \quad c_i, b_i\in\mathbb{R} .
	 \]
	Then, Lemma~2 allows us to obtain the general solution to the transformed equation \eqref{ztt}.

	\section{Solutions of the equations of acoustics and gas dynamics}
	
	The Riemann–Euler equation in the form (\ref{utt=}) is transformed, by a change of independent variable, into a one-dimensional equation of acoustics of inhomogeneous medium
		\begin{equation} \label{Ac}
		v_{tt} = f(y) v_{yy}.
	\end{equation}
		Indeed, introduce the function \(v(t, y(x)) = u(t, x)\) and compute the derivatives:
		\[
	u_{tt} = v_{tt}, \qquad u_x = v_y y^{\prime}, \qquad
	u_{xx} = v_{yy}(y^{\prime})^2 + v_y y^{\prime\prime}.
	\]
		Substituting these expressions into (\ref{utt=}), we obtain the equation
	\[
	v_{tt} = (y^{\prime})^2 v_{yy} + (y^{\prime\prime} + G y^{\prime}) v_y.
	\]
	Suppose that \(y(x)\) satisfies the equation
		\[
	y^{\prime\prime} + G y^{\prime}=0.
	\]
		Then \(y\) and \(y^{\prime}\) are given by 
	\begin{equation} \label{y}
		y' = \exp\left(-\int G dx\right), \qquad y = \int \exp\left(-\int G dx\right) dx
	\end{equation}
		Thus, the equation for the function \(v\) takes the form
		\begin{equation} \label{v_tt}
		v_{tt} = e^{-2\int G dx} v_{yy} .
	\end{equation}
	
	To make the coefficient in (\ref{v_tt}) an explicit function of \(y\), we need to express \(x\) from the second formula (\ref{y}) and substitute it into (\ref{v_tt}). This can be done explicitly in special cases.
	
	Let us give a classic example. Let \(G = n/x\), then the change of variables \(y = \frac{x^{1-n}}{1-n}\) maps the equation
		\begin{equation} \label{u_tt=}
		u_{tt} = u_{xx} + \frac{n}{x} u_x
	\end{equation}
	to the equation
	\begin{equation}
		v_{tt} = a y^{\frac{2n}{n-1}} v_{yy}, \qquad a = (1-n)^{\frac{2n}{n-1}}.
	\end{equation}
	By applying a scaling transformation, the constant 	$a$  can be set equal to 1.
		As noted earlier, solutions of equation \eqref{u_tt=} for even $n$ are expressed in terms of two arbitrary functions and their derivatives.
		Hence, equation (3.5) also possesses this property. These solutions were found by Euler \cite{Euler}.
	
	Let us now consider the equation
		\begin{equation} \label{utt=a}
		u_{tt} = u_{xx} + \frac{(n+2)x^{n+1} - a n}{x^{n+2} + a x} u_x,
	\end{equation}
	where \(a\) is an arbitrary constant, \(n\) is an even number.

	As previously described, solutions of equation (\ref{utt=a}) can be derived from those of equation (\ref{u_tt=}) by means of a transformation:
		\[
	u \longrightarrow \frac{u_x}{a x^{-n} + x}.
	\]

		Solutions of equation (\ref{u_tt=}) for even \(n = 2m > 0\) are given by the formula
\[ 	u = \left( \frac{1}{x} \frac{d}{dx} \right)^n (u_0),
	\]
	where \(u_0 = T(t+x) + X(x-t)\), \(T, X\) are arbitrary smooth functions. Then the corresponding solutions of equation (\ref{utt=a}) for \(n = 2, 4, 6\) have the form:
	\begin{align} \label{u246}
		u &= \frac{1}{x^3 + a} \left( V - U + x (U' + V') \right), \notag\\
		u &= \frac{1}{x^5 + a} \left( 3(V - U) - 3x(U' + V') + x^2(U'' - V'') \right), \\
		u &= \frac{1}{x^7 + a} \left( 15(V - U) + 15x(U' + V') - 6x^2(U'' - V'') + x^3(U''' + V''') \right), \notag
	\end{align}
	where \(U = T'\), \(V = X'\).
	
It is easy to see that the change of variable
\begin{equation} \label{y=}
	y = \frac{c}{x^{n+1} + a}, \quad c \in \mathbb{R}
\end{equation}
transforms equation \eqref{utt=a} into
\begin{equation} \label{vtt=n}
	v_{tt} = \left( \frac{n+1}{c}\right)^2 y^{\frac{2n+4}{n+1}} 
	\left( c - ay \right)^{\frac{2n}{n+1}} v_{yy}.
\end{equation}
According to \eqref{y=}, we express
\begin{equation} \label{x=}
	x = \left( \frac{(c - ay)^{\frac{1}{n+1}}}{y} \right). 
\end{equation}
Thus, solutions of equation \eqref{vtt=n} for \(n = 2, 4, 6\) are obtained from solutions \eqref{u246} of equation \eqref{utt=a} by substitution \eqref{x=}.

Now, let us transition from the acoustic equation (\ref{vtt=n}) to the gas dynamics equations in Lagrangian variables. To achieve this, we first apply the Legendre transformation.
\[
x = v_y, \quad \tau = v_t, \quad w = tv_y + yv_t - v.
\]
As a result, equation \eqref{vtt=n} transforms into
\[
w_{\tau \tau} = \left( \frac{c}{n+1} \right)^2 (c - a w_x)^{\frac{-2n}{n+1}} w_x^{-\frac{2n+4}{n+1}}w_{xx}.
\]
Let us rewrite this equation in divergent form:
\[
w_{\tau \tau} = \frac{\partial}{\partial x} \left(
\frac{  a^2 w_x^2(n+1)^2 + ca(1-n^2)w_x  + c^2 (1-n) }{c(n+3)(n^2-1)w_x^{\frac{n+3}{n+1} }(c-aw_x)^{\frac{n-1}{n+1}}} \right). 
\]
Denote \(w_x\) by \(\rho\), so  the last equation becomes 
\begin{equation} \label{wPr}
	w_{\tau \tau} = -\frac{\partial p(\rho)}{\partial x}, 
\end{equation}
where the function \(p\) is given by:
\begin{equation} \label{p}
	p = \frac{c^2(n-1)\rho^2 + ca(n^2-1)\rho - a^2(n+1)^2}{c(n+3)(n^2 - 1)} (c\rho - a)^{\frac{1 - n}{n+1}}. 
\end{equation}
As known \cite{Mises, CourantFriedrichs}, equation of the form \eqref{wPr} describes one-dimensional isentropic flow in Lagrangian variables. Knowing solutions of equation \eqref{vtt=n} and using the Legendre transformation, we can find solutions of equation \eqref{wPr}.

Let us now consider gas dynamics equations in Eulerian variables:
\begin{equation} \label{GAS}
	\rho_t + (u \rho)_x = 0, \qquad u_t + u u_x + \frac{p_x}{\rho} = 0, 
\end{equation}
where \(u\) is velocity, \(\rho\) is density, and \(p(\rho)\) is pressure.
Denote by \(Z\) the speed of sound, equal to \(\sqrt{p'(\rho)}\), and introduce Riemann invariants
\begin{equation} \label{rs}
	r = u + \int \frac{Z(\rho) \, d\rho}{\rho} \ , \quad s = u -\int \frac{Z(\rho) \, d\rho}{\rho}\ , 
\end{equation}
rewriting equations \eqref{GAS} in terms of invariants \cite{Riem, Roz}
\begin{equation}
	r_t + (u + Z(\rho)) r_x = 0, \qquad s_t + (u - Z(\rho)) s_x = 0. 
\end{equation}

If we express \(u\) and \(\rho\) in terms of the Riemann invariants \(r, s\) from relations \eqref{rs}, we obtain the equations
\begin{equation*}
	r_t + \left(\frac{r + s}{2}  + K(r - s)\right) r_x = 0, 
	\qquad s_t + \left(\frac{r + s}{2} - K(r - s)\right) s_x = 0. 
\end{equation*}
As Riemann noted \cite{Riem,Roz}, this system has solutions in implicit form
\begin{equation} \label{x-t}
	x - \frac{r + s}{2} t - K(r - s) t =w_r, \quad x - \frac{r + s}{2} t +
	K(r - s) t = w_s,
\end{equation}
where the function \(w(r, s)\) satisfies
\begin{equation} \label{wrs=}
	w_{rs} + \frac{2K^{\prime} - 1}{4K} (w_r - w_s) = 0. 
\end{equation}

Now assume the pressure is given by
\begin{equation} \label{p=}
	p = \frac{ (\rho - a)^{\frac{n+3}{n+1}}}{(n+1)(n+3)} + 
	\frac{a (\rho - a)^{\frac{2}{n+1}}}{n+1} - 
	\frac{a^2 (\rho - a)^{\frac{1-n}{n+1}}}{n^2 - 1}.
\end{equation}
Note that this function coincides with \eqref{p} when \(c = 1\). Moreover, for \(a = 0\) it describes the equation of state of a polytropic gas.
Consequently, the Riemann invariants according to \eqref{rs} become
\[ r = u + (\rho - a)^{\frac{1}{n}}, \quad s = u - (\rho - a)^{\frac{1}{n}}. \] 
 It follows that
\[ u = \frac{r + s}{2}, \qquad
\rho = \left( \frac{r - s}{2} \right)^{n+1} + a, \] 
\[ K = \frac{1}{n+1} \left( a\left( \frac{r-s}{2} \right)^{-n} + \frac{r-s}{2} \right). \]  
It is easy to see that, in this case, equation \eqref{wrs=} reduces to the EPD equation
\begin{equation} \label{las}
	 w_{rs} - \frac{n}{2(r-s)} (w_r - w_s) = 0.
\end{equation}
As noted earlier, if \(n\) is an even number, the solution of this equation is expressed in terms of arbitrary functions. This particularly follows from the results of the previous section. Assuming \(n=2\), the general solution of equation \eqref{las} takes the form
\[ w = \frac{T(r) + X(s)}{r-s}, \] 
where \(T, X\) are arbitrary functions. 
In this case, formulas \eqref{x-t} provide implicit solutions to the gas dynamics system expressed via  Riemann invariants.

\section{Conclusion}

This work finds general solutions of one-dimensional non-stationary equations of acoustics and gas dynamics for a special equation of state. 
For this purpose, we relied on Euler-Darboux transformations and Riemann's method.
 This approach extends to other equations. 
For example, consider the simplest system in plasticity theory \cite{Novatsky}
\[ q_t = v_x, \quad v_t = a^2(q) q_x. \] 
In Riemann invariants it takes the form
\[ r_t + b(r-s) r_x = 0, \qquad s_t = b(r-s) s_x. \]
The implicit solution:
\[ x - b t = w_r, \qquad x + b t = w_s \]
is obtained by Riemann's method. Here the function \(w\) satisfies
\[ w_{rs} + \frac{b'}{2b} (w_r - w_s) = 0. \]  
This equation admits general solutions when \(b=(r-s)^{2n}\) for integer \(n\).  
The work \cite{Tsarev} proposes an integration method for semi-Hamiltonian hyperbolic systems, generalizing Riemann's approach. The works \cite{KapSib} and monograph \cite{KaptsovBook} describe a method for integrating nonlinear hyperbolic systems using "higher-order" characteristic invariants.

\end{document}